\documentclass[twocolumn,letterpaper]{emulateapj}
\usepackage{psfig}
\usepackage{epstopdf}
\usepackage{natbib}
\submitted{Draft version: \today}

\def\alwaysmath#1{\ifmmode{#1}\else{$#1$}\fi}

\shortauthors{Gon\c{c}alves et al.}
\shorttitle{Mass flux in the green valley}

\begin{document}

\title{Quenching star formation at intermediate redshifts: downsizing of the mass flux density in the green valley} \author{\sc Thiago S. Gon\c{c}alves,\altaffilmark{1,}\altaffilmark{2} D. Christopher Martin,\altaffilmark{1} Kar\' in Men\'endez-Delmestre,\altaffilmark{2} Ted K. Wyder,\altaffilmark{1} Anton Koekemoer\altaffilmark{3}}

\altaffiltext{1}{California Institute of Technology, MC 278-17, Pasadena, CA 91125}

\altaffiltext{2}{Observat\'orio do Valongo, Universidade Federal do Rio de Janeiro}

\altaffiltext{3}{Space Telescope Science Institute}

\email{tsg@astro.ufrj.br}

\begin{abstract}

The bimodality in galaxy properties has been observed at low and high redshift, with a clear distinction between star-forming galaxies in the blue cloud and passively evolving objects in the red sequence; the absence of galaxies with intermediate properties indicates that the quenching of star formation and subsequent transition between populations must happen rapidly. In this paper, we present a study of over 100 transiting galaxies in the so-called ``green valley'' at intermediate redshifts ($z\sim 0.8$). By using very deep spectroscopy with the DEIMOS instrument at the Keck telescope we are able to infer the star formation histories of these objects and measure the stellar mass flux density transiting from the blue cloud to the red sequence when the universe was half its current age. Our results indicate that the process happened more rapidly and for more massive galaxies in the past, suggesting a top-down scenario in which the massive end of the red sequence is forming first. This represent another aspect of downsizing, with the mass flux density moving towards smaller galaxies in recent times.

\end{abstract}


%
% Section 1
%

\section{Introduction}

The distinction between blue star-forming galaxies and red, passively evolving ones has been know for a long time. In his seminal classification article, \cite{Hubble1926} classified galaxies into two main groups, spirals and elllipticals. Although unintended as a timeline for secular evolution -- in Hubble's own words `The entire classification is purely empirical and without prejudice to theories of evolution' \citep{HubbleE.P.1927} -- ellipticals were referred to as `early-type galaxies' and spirals as `late types' \citep[see also][]{FortsonLucy2011}.

Our knowledge of stellar populations in galaxies has evolved considerably since then. Today we know that spiral galaxies are blue due to young stars, and that their current star formation rates are higher than in ellipticals, with a younger population, on average. \citet{Davis1976} were the first to show that clustering properties of galaxies depend on galaxy type, with ellipticals showing stronger clustering properties than spirals. We also know that elliptical galaxies in the local universe are preferentially found in environments where the galaxy number density is higher such as clusteres of galaxies \citep{Dressler1980}, which gives us important clues regarding the late evolution of galaxies. The same relation has been shown to extend to groups of galaxies \citep{Postman1984} and environments such as superclusters \citep{Giovanelli1986}. This leads to the conclusion that spiral galaxies {\it somehow} transform into ellipticals at later times, for example through secular evolution or mergers with other galaxies. However, the existing bimodality in the galaxy distribution, with a clear distinction between blue spirals and red spheroids, and its establishment still remain a puzzle.

Other works have correlated the bimodality in galaxy properties with galaxy colors. \cite{Visvanathan1977} were the first to show the correlation between colors and magnitudes of cluster galaxies in the local universe. The bimodality in galaxy colors and their correlations with galaxy morphology was first noted by \cite{Takamiya1995}. More recently, the color bimodality has been observed at low redshifts ($z\sim 0.1$) with the Sloan Digital Sky Survey (SDSS; \citealt{Strateva2001,Baldry2004}). Later, many authors have shown this bimodality extends to earlier times, at least out to $z\sim 1.0$ \citep{Koo1996, Balogh1998, Im2002, Bell2004, Weiner2005, Willmer2006}. As is the case for galaxy morphologies, several recent works have shown that galaxy colors are correlated with environment density \citep[e.g.,][]{Blanton2005,Cooper2006}. If the color distribution of galaxies can be described as two distinctive peaks, then one can denominate the minimum at intermediate color values the ``green valley''. The question then arises: why does such a minimum exist, instead of a homogeneous distribution across the color-magnitude diagram?

In order to study this intermediate population, one needs to determine proper selection criteria. \cite{Wyder2007} have measured number densities in the color-magnitude diagram at low redshift ($z\sim 0.1$) and found that the GALEX ${\rm NUV} - r$ color represents more efficient criterium to select green valley galaxies. Ultraviolet emission originates from recent (over the last tens of millions of years) star-forming regions in the galaxy, while the $r$ band is more sensitive to the bulk of stellar mass, formed over the course of the galaxy's history. By subtracting the two, one can clearly distinguish two populations, one actively forming stars and the other older, more passive, with a dynamic range of about six magnitudes in color. Nevertheless, we need to be careful when defining transition galaxies; the percentage of star-forming, obscured galaxies is high. These are typically galaxies with high ratio of current to past star formation, but which present redder colors due to dust obscuration \citep[see also section \ref{sec:extinction}]{Martin2007,Salim2009}.

The low number density (or number deficit) of galaxies in the green valley indicates that the transition between both groups occurs rapidly. A number of works attempt to explain why there is such a rapid evolution. \citet{Menci2005} postulate that supernova winds and outflows can play an important part, driving out gas that fuels star formation in the galaxy. \citet{DiMatteo2005}, on the other hand, have produced a hydrodynamical simulation of a major merger event which shows that, after a period of brief increase of star formation, the supermassive black hole created in the center of the merger remnant drives strong outflow winds that rapidly quench star formation. \citet{Nandra2007} and \citet{Schawinski2009} provide observational support to this hypothesis, finding a large number of active galactic nuclei (AGN) in the green valley and concluding that feedback from such objects might be somehow quenching the star formation process. This is further supported by \citet{Coil2008}, who note that the coadded spectra of green valley galaxies at $z\sim 1$ are distinct from coadded spectra of blue and red galaxies, with line ratios that show increased AGN activity. \citet{Peng2010} go further and state that quenching is a combination of a mass-dependent quenching mechanism, proportional to the galaxy's star formation rate, and the effect of the hierarchical assembly of dark matter haloes. Nevertheless, \citet{Mendez2011} note that the merger fraction in the green valley is low, from the quantitative morphological analysis of $0.4<z<1.2$ optically selected transition galaxies. The authors then conclude that mild external processes (such as galaxy harassment) or secular evolution are the dominant factors at these redshifts.

This rationale assumes, however, that galaxies move in one direction only in the color-magnitude diagram, from blue to red. Nevertheless, there are several cases in the literature that indicate ongoing and/or recent gas accretion, in addition to comparatively high star formation rates in elliptical galaxies \citep[e.g.,][]{Morganti2006,Kaviraj2008,Catinella2010,Schiminovich2010}. Although these cases are interesting, they are relatively rare, and one can in general assume that galaxies with intermediate colors are currently in the process of quenching their star formation.

To infer how rapidly galaxies are moving across the green valley, \citet{Martin2007} (hereafter Paper I) have used spectroscopic features in green valley galaxies to obtain information on their star formation histories (see section \ref{sec:method} for details). Along with measurements of typical galaxy masses and number densities in the color-magnitude diagram, the authors have been able to determine the mass flux across the green valley at low redshifts ($z\sim 0.1$). The measured value of $\dot\rho=0.033$ M$_\odot$ yr$^{-1}$ Mpc$^{-3}$ agrees remarkably well with expectations from the growth of the red sequence and the depletion of galaxies in the blue sequence at such redshifts \citep{Bell2004,Faber2007}, despite the shortcomings of the simple model assumed. Furthermore, Paper I presents evidence for an increase in number density of AGN in the green valley (from measured [OIII] luminosities), supporting the aforementioned studies that relate AGN activity with the quenching of star formation (although the correlation between [OIII] luminosities and quenching timescales is not unequivocal).

In this work, we apply the methods introduced in paper I to a sample of galaxies at higher redshift. This will allow us to measure the mass flux in the green valley at intermediate redshifts, comparing with results found for galaxies in the low-redshift universe. Although the technique is simple, the data acquisition process proves challenging, since it requires reliable measurements of absorption features in galaxies at redshifts of ($z\sim 0.8$). We have set out to obtain the required data, and the work presented here includes the deepest spectra of green valley galaxies to date.

This paper is organized as follows. In section \ref{sec:method}, we describe in detail the methodology used to infer star formation histories of galaxies, including modelling of stellar populations. In section \ref{sec:gvobs} we describe the observations and data reduction used to produce the spectra which were then used to measure star formation histories and quenching timescales, in addition to ancillary data used to measure number densities in the color magnitude diagram, luminosity functions and extinction correction. Section \ref{sec:gvresults} shows our results, including the measured mass flux density at $z\sim 0.8$ and in section \ref{sec:gvdiscussion} we discuss those results in light of galaxy evolution models. We summarize our findings in section \ref{sec:gvsummary}. Throughout this paper, we use AB magnitudes, and assume standard cosmology, with $H_0 = 70$ km s$^{−1}$ Mpc$^{−1}$, $\Omega_m = 0.30$, and $\Omega_\Lambda = 0.70$.

\section{Methodology}\label{sec:method}

The method to study the mass flux in the green valley has been introduced by \cite{Martin2007}. We summarize here the description presented in that work.

\subsection{The mass flux density in the color magnitude diagram}

In order to measure the mass flux for a given color in the color-magnitude diagram, $\dot{M}(r,y)$, where $y$ is the $NUV-r$ color of the galaxy, we can assume that

\begin{equation}
\dot{M}(r,y) = <M(r,y)> \times \frac{dy}{dt}, \label{eq:flux_gv}
\end{equation}
where $<M(r,y)>$ is the average mass of a galaxy in that color-magnitude bin, and $dy/dt$ is how fast galaxies are moving through the same bin. In practice, we measure average values in a two-dimensional bin. Dividing the above equation by the comoving density probed, we calculate

\begin{equation}
\dot{\rho}(r,y) = \Phi(r,y)<M(r,y)>\frac{dy}{dt}. \label{eq:flux_density_gv}
\end{equation}
In this case, the comoving mass density $\rho(r,y)$ is simply the comoving number density ($\Phi(r,y)$) multiplied by the typical galaxy mass in that bin ($<M(r,y)>$). What we propose to measure is the mass flux density $\dot\rho$. Since we can constrain $\Phi(r,y)$ and $<M(r,y)>$ independently from the star formation histories, we are only left with the task of measuring the color evolution rate for a given galaxy or bin.

\subsection{Star formation histories}

In order to measure the color evolution rates, we make some simplifying assumptions:

\begin{enumerate}

\item Galaxies only move towards redder colors, i.e. we do not consider starbursting red galaxies moving downward in the color-magnitude diagram;

\item The star formation histories in all galaxies are described by a constant star formation rate for the first few Gyr (approximately 6 Gyr) followed by a period of exponentially declining star formation rates:

\begin{equation}
{\rm SFR}(t) = \left\{ \begin{array}{rl}
 {\rm SFR}(t=0), & t<t_0 \\
 {\rm SFR}(t=0)e^{-\gamma t}, &t>t_0
       \end{array} \right.
\label{eq:sfh}
\end{equation}

\end{enumerate}
We discuss the implications of this model in Section \ref{sec:sfh}.

In order to measure the exponential index $\gamma$, we apply the same methodology described in \citet{Kauffmann2003} to measure the rest-frame 4000 {\AA} break and the equivalent width of $H_\delta$ absorption. The first is created by the accumulation of a large number of ionized metal absorption lines shortward of 4000 {\AA} \citep{BruzualA.1983}. The elements responsible for these lines are multiply ionized in hot stars, resulting in smaller opacities in the integrated spectra of younger galaxies, and thus the break correlates with the age of the stellar population. The latter is mainly present in the stellar photosphere of smaller early-type stars (mostly A stars), which dominate $0.1-1.0$ Gyr after a starburst event (and after the O and B stars evolved off the main sequence).

In this work, we use the same definitions as in Paper I, in order to maintain uniformity across different redshifts. The first index, $D_n(4000)$, is defined as the ratio of the average flux density $F_\nu$ in the bands 3850-3950 and 4000-4100 \AA, following the definition of \citet{Balogh1999}. The narrow bands ensure that the ratio is weakly dependent on flux calibrations and other broad band effects that may arise from data reduction. For the latter, $H_{\delta,A}$ is the absorption equivalent width; the continuum is defined by fitting a straight line through the average flux density between 4041.60 and 4079.75 \AA, on one end, and 4128.50 and 4161.00 \AA, on the other. The equivalent width, then, is given simply by

\begin{equation}
H_{\delta,A} = \sum_{\lambda=4083.5}^{4122.25}\left(1-\frac{F_\nu}{F_{\nu,cont.}}\right).\label{eq:hda}
\end{equation}

\citet{Kauffmann2003} showed that these indices trace a well defined region in a $Dn(4000)$ vs. $H_{\delta,A}$ diagram. Furthermore, different star formation history tracks -- for instance a single starburst event versus a continuous star formation rate -- trace distinct regions within this diagram. We therefore use our measured spectral indices as defined above to distinguish between different star formation histories, in our case as given by different $\gamma$ values.

\begin{figure}[htp]

\includegraphics[width=\linewidth]{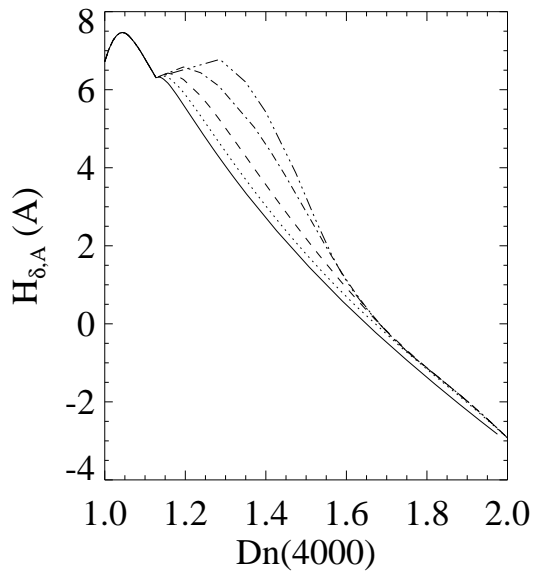}

\caption[$D_n(4000)$ vs. $H_{\delta,A}$ for different models]{Evolution of $D_n(4000)$ vs. $H_{\delta,A}$ for models with different star formation histories. Solid, dotted, dashed, dotted-dashed and triple-dotted-dashed lines (from left to right) represent $\gamma$'s equal to 0.5, 1.0, 2.0, 5.0 and 20.0 Gyr$^{-1}$, respectively. Therefore, models to the right are quenching star formation more rapidly.}\label{fig:dn_hd_model}

\end{figure}

In Figure \ref{fig:dn_hd_model}, we show 5 tracks on the $Dn(4000)$ vs. $H_{\delta,A}$ diagram given by 5 distinct $\gamma$ values (0.5, 1.0, 2.0, 5.0 and 20.0 Gyr$^{-1}$). These models were produced with the \citet{Bruzual2003} models -- with \citet{Chabrier2003} initial mass functions, Padova 1994 stellar evolutionary tracks and solar metallicities -- by varying ages throughout star formation histories as described in equation \ref{eq:sfh}. The range of values for $t_0$ is limited to $t_0<t_z$, so that the oldest models are always below the universe age at the highest redshift we measure. This does not have a major impact on our results, since $t_0$ is typically larger than the time when $D_n(4000)$ and $H_{\delta,A}$ have stabilized. We notice very distinct tracks, with the strongly quenched models (higher $\gamma$ presenting higher $H_{\delta,A}$ values).

\begin{figure}[htp]

\includegraphics[width=\linewidth]{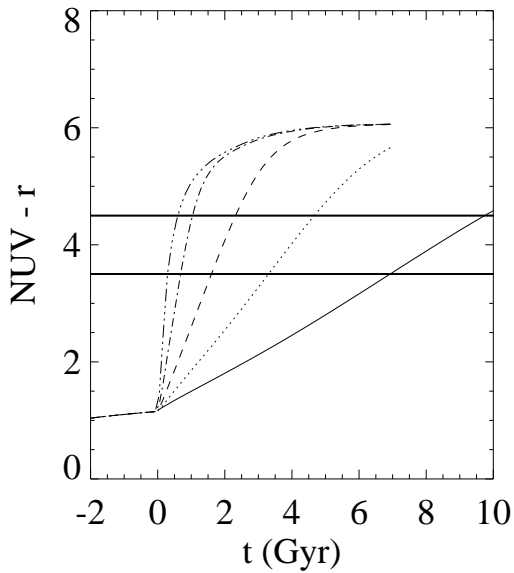}

\caption[Time evolution of NUV-$r$ colors in models]{${\rm NUV}-r$ colors of model galaxies as a function of time after the onset of exponential decline in SFR. Line types are the same as in Figure \ref{fig:dn_hd_model}. The thick horizontal lines indicate the threshold we have used to determine the green valley. $dy/dt$ will then simply be the width of the magnitude band in colors divided by the time take for a galaxy to cross both thresholds.}\label{fig:age_nuv_r_model}

\end{figure}

Finally, the measurement of $dy/dt$ is given by the width of the color range in magnitudes used to determine the green valley divided by the duration of time when the galaxy is within those boundaries. We illustrate this in Figure \ref{fig:age_nuv_r_model}, where we plot the ${\rm NUV}-r$ color as a function of time after the initial period of constant star formation, with the green valley limits indicated by horizontal lines. As expected, models with higher values of gamma (i.e. that are quenching their star formation more rapidly) change colors faster. The timescales to cross the green valley in these models varies between 0.26 and 2.7 Gyr.

\section{Sample, observations and data processing}\label{sec:gvobs}

\subsection{Sample selection}

We have selected a sample of 163 green valley galaxies from surveys at intermediate redshift surveys. Preference has been given to galaxies in the Extended Groth Strip (EGS), which is the target of the ongoing All-wavelength Extended Groth Strip International Survey collaboration \citep[AEGIS;][]{Davis2007}. This field is optimum for galaxy evolution studies, since the AEGIS collaboration has produced large amount of ancillary multiwavelength data. In particular, a large subsample of galaxies has spectroscopically determined redshfits with the DEEP2 survey \citep{Davis2003,Newman2012}. However, in order to facilitate observations throughout the year, we have also used data from the Canada-France-Hawaii Telescope Legacy Survey (CFHTLS)\footnote{http://www.cfht.hawaii.edu/Science/CFHTLS/}, from all deep fields observed.

All galaxies observed in this work have CFHTLS photometry in all 5 bands ($u$,$g$,$r$,$i$,$z$). This is required to properly constrain SDSS r' band magnitudes in the galaxy's restframe, while simultaneously yielding accurate GALEX NUV restframe magnitudes. In order to match the magnitude limit of the spectroscopic sample in DEEP2, we have determined a threshold of $M_r\leq 24$ for all objects in our sample. The CFHTLS Deep Field D3 overlaps with the EGS, and in that case we match CFHTLS and DEEP2 sources. We preselect galaxies with redshifts between $0.55 \leq z \leq 0.9$; DEEP2 redshifts were used for the selection in the EGS field, CFHTLS photometric redshifts are used otherwise (approximately 50\% of our final sample).

The final sample of 163 observed galaxies was selected at random from the available targets, in an attempt to maximize object count in a given DEIMOS mask. Not all objects in a given area were available for observation in a single pointing, given the number density of green valley galaxies and the design constraints for slits in each mask. Preference was given to objects in the green valley after extinction correction, but no additional criteria are used in source selection.

\begin{figure}[htp]

\includegraphics[width=\linewidth]{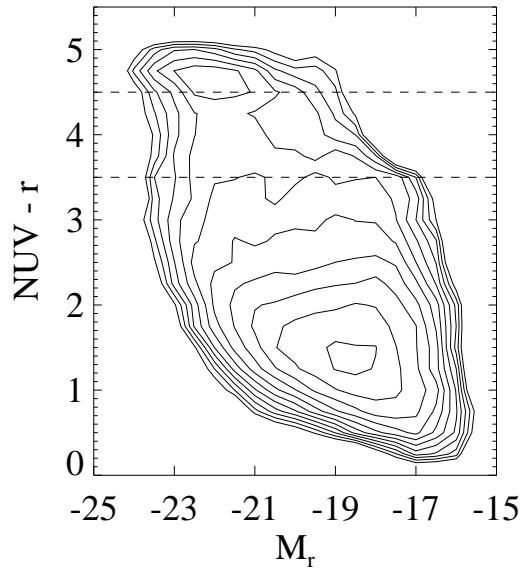}

\caption[Color-magnitude diagram of CFHTLS sources at intermediate redshifts]{Color-magnitude diagram of all CFHTLS sources between $0.55<z<0.9$ in logarithmic contours. Two distinct populations are easily distinguishable, with the ``green valley'' in-between. The dashed lines indicate the color selection criterium used in this work.}\label{fig:cmplot_cfhtls}

\end{figure}

All sources have been k-corrected to redshift $z=0$ using the Kcorrect code \citep[version 4\_2;][]{Blanton2007}, so that we could select them based on rest-frame NUV-r colors. In Figure \ref{fig:cmplot_cfhtls} we show the result of the k-correction processes, showing all galaxies below the completeness limits in CFHTLS-Deep. We also indicate the color selection criterium, showing the color limits for the green valley galaxies as dashed lines.

%In Figure \ref{fig:gv_hst} we show image stamps obtained with HST in EGS field with the ACS instrument in the $V$ and $I$ bands \citep{Davis2007,Lotz2008}. We note that although most portrayed galaxies are small spheroids, the presence of disks is significant ($\leq 40$\%). One of them is clearly an ongoing merger. A detailed morphological analysis is beyond the scope of this work; for an in-depth discussion on the morphology of green valley galaxies, see \citet{Mendez2011}.

\subsection{Observations and Data Reduction}

Our observations were undertaken with the DEIMOS instrument on the Keck II telescope \citep{Faber2003}. DEIMOS is a multi-object optical spectrograph, which allows up to nearly 100 objects per mask in a field of view of 16.7 arcmin x 5.0 arcmin. The number of GV galaxies per mask oscillated between 20-30 obj/mask; other galaxies in the aforementioned surveys were then used to complement observations. We used the 1200 mm$^{-1}$ grating for a resolution of $R\sim 5000$. A grating angle centered on 7500 {\AA} was chosen in order to cover all the necessary wavelength range needed to measure D$_n(4000)$ and H$_{\delta,A}$ at the redshift range of our sample. In preparing the masks, preference was given to objects in the extinction-corrected green valley (section \ref{sec:extinction}).

In order to achieve the required signal-to-noise level ($S/N \simeq 3$ per pixel, or $S/N\simeq 2$ for galaxies fainter than $r>23$) , we exposed for a total of up to 9 hours per mask, weather and sky availability permitting. This comprises the deepest spectra of intermediate redshift green valley galaxies taken to date. The observations are summarized in Table \ref{table:deimos_obs}.

Preliminary data reduction was performed with the DEEP2 pipeline\footnote{The pipeline is available at http://deep.berkeley.edu/spec2d/}. The pipeline is currently optimized for measuring redshifts in galaxies observed with the DEEP2 survey; therefore, we have used it to select individual spectra, rectify slits and subtract the sky background. Extraction to one-dimensional spectra, refinement of the wavelength solution through the use of sky lines and further data analysis have been done with custom IDL procedures. All redshifts have been remeasured to ensure correct wavelengths for measuring D$_n(4000)$ and H$_{\delta,A}$.

\begin{figure}[htp]

\includegraphics[width=\linewidth]{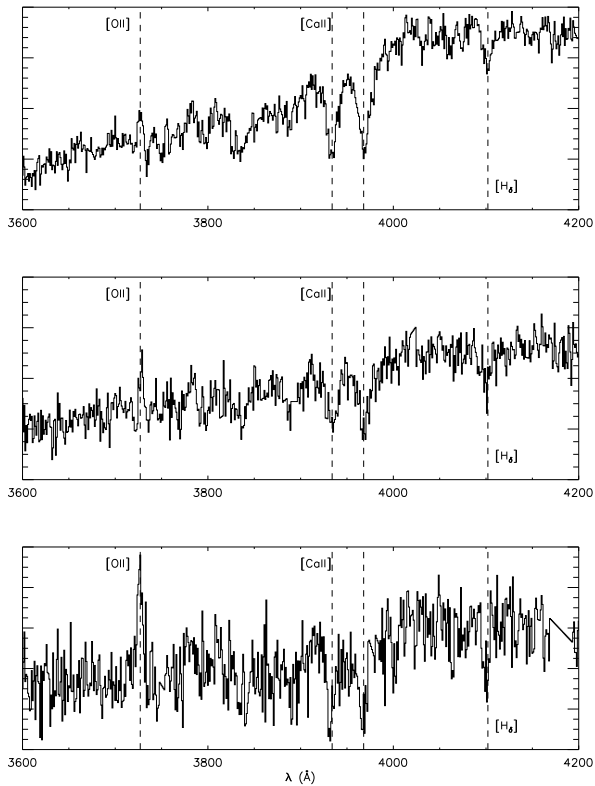}

\caption[Green valley DEIMOS spectra]{Spectra of three green valley galaxies plotted in the region of interest (between $3600 \AA<\lambda<4200 \AA$). Observed $r$ magnitudes are approximately 21.5, 22.5 and 23.5, respectively. Spectra have been re-sampled to match the model resolution of 1 {\AA} per pixel. We also highlight the position of [OII], CaII H \& K and H$_\delta$ lines. In all cases the calcium and H$_\delta$ lines are clearly distinguishable.}\label{fig:gv_spec}

\end{figure}

When measuring D$_n(4000)$ and H$_{\delta,A}$ values for individual galaxies, we have resampled the spectra so that the wavelength resolution per pixel is the same as the \citet{Bruzual2003} models, i.e. 1 \AA. In all cases we mask out regions contaminated by sky lines. We show the resulting spectra in Figure \ref{fig:gv_spec}.

\begin{table}%{lcccccccc}
\begin{center}
%\tablewidth\linewidth
%\tabletypesize{\scriptsize}
\caption{Summary of DEIMOS Observations} 
\begin{tabular}{c c c c c}\hline
%\tablehead{
Mask & Field & Observing & Total integration & Number of \\
  & & date (UT) & time & GV galaxies\\
%  }
%\startdata
\hline
AG0801 & EGS & May 2008 & 32400 & 45\\
AG0803 & EGS & May 2008 & 25200 & 48\\
DZLE01 & DEEP2 Field 2 & May 2008 & 3600 & 18\\
D401 & CFHTLS D4 & August 2008 & 16200 & 37\\
XMM02 & XMM-LSS & August 2008 & 7200 & 42\\
AG0901 & EGS & April 2009 & 28800 & 33\\
AG0902 & EGS & April 2009 & 28800 & 30\\
\hline
\label{table:deimos_obs}
%\enddata
\end{tabular}
\end{center}
%\tablenotetext{a}{in mas}
%\tablenotetext{b}{UV half-light radius from HST data}
\end{table}

\subsection{Number densities and the luminosity funcion}\label{sec:density}

In order to determine the number density per bin in the color-magnitude diagram, we have used only CFHTLS photometric data, including photometric redshifts when spectroscopic data are not available. These redshifts are precise enough for this exercise, and the increase in sample size and area greatly improves the statistics. Furthermore, since the CFHTLS includes 4 different fields in distinct regions of the sky (D1, D2, D3, D4), the influence of cosmic variance is minimized.

In determining number densities, we have used the $1/V_{\rm max}$ method \citep{Schmidt1968}. This takes into account the magnitude limits of the survey and the potential of missing low-luminosity galaxies. In that sense, we perform a k-correction of every source in the survey between the limiting redshifts ($0.55<z<0.9$) to determine the maximum distance at which we would be able to detect it (below the $z=0.9$ cutoff), taking into account all 5 bands used to select sources. We denote the maximum redshift as

\begin{equation}
z_{\max} = {\rm min}(z_{\rm CFHTLS, max}, 0.9).
\end{equation}
$z_{\rm min}$ is simply 0.55, since there is no brightest magnitude cutoff. The maximum volume for each galaxy will be

\begin{equation}
V_{\rm max} = \frac{A}{3}\left(\frac{\pi}{180}\right)^2\left(\frac{D_L(z_{\rm max})^3}{(1+z_{\rm max})^3} - \frac{D_L(z_{\rm min})^3}{(1+z_{\rm min})^3}\right),
\end{equation}
where $A$ is the angular area in the sky in square degrees occupied by the four fields of the survey and $D_L(z)$ is the luminosity distance to redsfhit $z$.

\begin{figure}[htp]

\includegraphics[width=\linewidth]{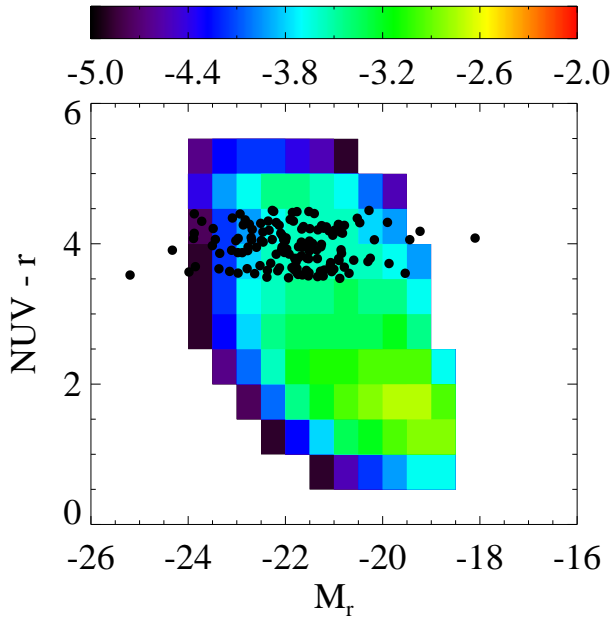}

\caption[Number density of galaxies in the color-magnitude diagram]{Diagram of number density of galaxies in each two-dimensional bin in the color-magnitude diagram, in \#/Mpc$^{-3}$. Two populations are distinguishable: the red sequence at ${\rm NUV} - r \sim 5$ and the blue sequence at ${\rm NUV} - r \sim 2$. Green valley galaxies used in this study (black points) fall in-between these two populations, at $3.5 < {\rm NUV} - r < 4.5$.}\label{fig:density}

\end{figure}

\begin{figure}[htp]

\includegraphics[width=\linewidth]{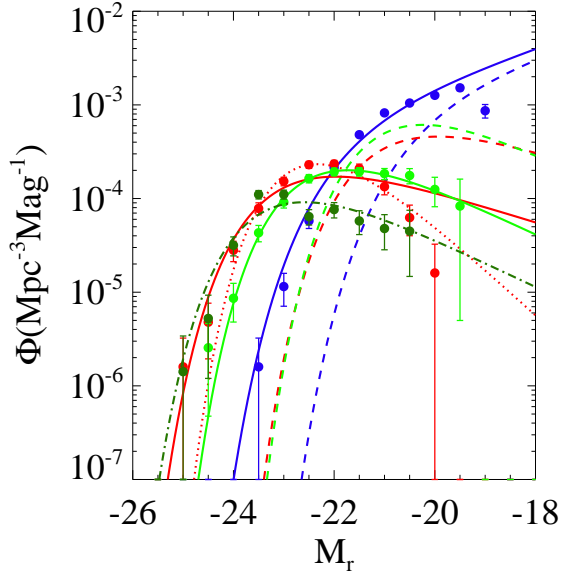}

\caption[Luminosity functions of different samples]{Luminosity functions at $z\sim0.8$ measured from the CFHTLS Deep survey. Blue represents the blue sequence (${\rm NUV}-r=1.75$), red, the red sequence(${\rm NUV}-r=4.75$), and green, the green valley (${\rm NUV}-r=3.75$). Solid lines represent the fit for a fixed faint-end slope equal to values found at low redshift. Additionally, the red dotted line represents the Schechter function fit with the faint-end slope as a free parameter. The dash-dotted dark-green line and symbols represent the extinction-corrected luminosity function for green-valley galaxies. Error bars are given by Poisson statistics only; cosmic variance errors should be minimal, since we use all four deep fields in CFHTLS to compute the luminosity functions. The dashed lines indicate luminosity functions measured at $z=0.1$ by \citet{Wyder2007}, for comparison. All functions at higher redshift are shifted towards brighter magnitudes. The blue sequence shows similar number densities, while red and green galaxies are rarer than in the local universe.}\label{fig:lumfunction}

\end{figure}

Finally, the number density in the color-magnitude diagram is
 \begin{equation}
 \Phi(M_r,y) = \frac{1}{\Delta M_r \Delta y} \sum\frac{1}{V_{\rm max}},
 \end{equation}
in units of Mpc$^{-3}$ mag$^{-2}$, where $\Delta M_r$ and $\Delta y$ are the magnitude and color bin size, respectively.. We show our results in Figure \ref{fig:density}. The colored regions indicate number densities as measured from the CFHTLS sample, while the black dots are the objects targeted in this work. Two regions of higher number density stand out, evidence of the color bimodality still present at these higher redshifts, similar to the results of \citet{Willmer2006}. However, a comparison with similar work at redshift $z\sim 0.1$ \citep{Wyder2007} shows that the number densities in the red sequence are smaller, showing there has been a significant growth since $z\sim 1$ \citep[][see also Figures \ref{fig:cmplot_cfhtls} and \ref{fig:lumfunction} and Figure 7 in \citet{Wyder2007}]{Faber2007}. We discuss this in more detail in section \ref{sec:gvdiscussion}.
 
We generate luminosity functions by multiplying the number densities by the color bin size $\Delta y$. In Figure \ref{fig:lumfunction}, we show the resulting luminosity functions for three different color bins (${\rm NUV}-r =1.75, 3.75, 4.75$), and compare those with the same functions at $z=0.1$, from \citet{Wyder2007}. Although galaxies have been k-corrected to different redshifts ($z=0$ for our sample and $z=0.1$ for \citet{Wyder2007}), the typical shift in magnitude is $M_r=0.1$, and thus this difference has no impact on our results.

Assuming the dynamic range is insufficient to constrain the faint-end slope of the luminosity funcion, we fix it at the same values as for the luminosity function at low redshift, following the strategy of \citet{Bell2004}, \citet{Willmer2006} and \citet{Faber2007}. As Figure \ref{fig:lumfunction} shows, this is satisfactory for the blue sequence and the green valley; the luminosity function of the red sequence, however, clearly shows a deficit of faint galaxies when compared to its low-redshift counterpart. In this case we also fit a Schechter luminosity function with the faint-end slope as a free parameter. We discuss this evolution in section \ref{sec:gvdiscussion}.

We see that (1) green valley and red sequence number densities are smaller than in the local universe, while number densities are similar for the blue sequence, and (2) all luminosity functions are shifted towards higher luminosities - a fact that has been extensively observed at high redshift and has been cited numerously as evidence for another mode of downsizing \citep[e.g][]{Bell2004,Bundy2006,Faber2007}. It should be noted that the red peak in the color-magnitude occurs at slightly different colors at low and high redsfhit, with the red sequence being slightly bluer at $z\sim 0.8$. However, the variation between the ${\rm NUV} - r=4.75$ and ${\rm NUV} - r=5.75$ luminosity functions at $z=0.1$ are small, and chosing a different color range would not change our results. In Table \ref{table:lumfunction} we present the resulting parameters from a Schechter function fit in each case, as well as the low-redshift values presented in \citet{Wyder2007}. The luminosity function is described as follows:

\begin{equation}
\Phi(M)=0.4\ln(10)\Phi^*10^{-0.4(M-M^*)(\alpha+1)}\exp[-10^{-0.4(M-M^*)}],
\end{equation}
where $M^*$ and $\Phi^*$ are the free parameters in the fit.

%\begin{deluxetable*}{lccc}
\begin{table*}
\begin{center}
%\tablecolumns{4}
%\tablewidth\linewidth
%\tabletypesize{\scriptsize}
%\tablecaption{Schechter function parameters} 
\caption{Schechter function parameters} 
\begin{tabular}{l c c c}\hline
%\tablehead{
%& \colhead{a} &&\\
%& \colhead{a} &&\\
%\colhead{Sample} & \colhead{Log $\Phi^*$} & \colhead{$M_*$} & \colhead{$\alpha$}\\
%  & \colhead{(Mpc$^{-3}$ Mag$^{-1}$)} &  &\\
Sample & Log $\Phi^*$ & $M_*$ & $\alpha$\\
%  }
%\startdata
\hline
Blue sequence (${\rm NUV}-r = 1.75$) & $-3.05\pm 0.05$ & $-21.73\pm 0.05$ & $-1.465$\\
Red sequence (${\rm NUV}-r = 4.75$) & $-3.39\pm 0.05$ & $-22.90\pm 0.06$ & $-0.579$\\
Red sequence (free $\alpha$) & $-3.17\pm 0.05$ & $-22.10\pm 0.11$ & $0.24\pm 0.14$\\
Green valley (${\rm NUV}-r = 3.75$) & $-3.26\pm 0.05$ & $-22.20\pm 0.06$ & $-0.357$\\
Green valley (extinction-corrected) & $-3.60\pm 0.07$ & $-23.07\pm 0.08$ & $-0.357$\\
Blue sequence ($z=0.1$) & $-2.871$ & $-20.331$ & $-1.465$\\
Red sequence ($z=0.1$) & $-2.962$ & $-20.874$ & $-0.579$\\
Green valley ($z=0.1$) & $-2.775$ & $-20.711$ & $-0.357$\\
\hline
\label{table:lumfunction}
%\enddata
\end{tabular}
\end{center}
%\end{deluxetable*}
\end{table*}

\subsection{Extinction correction}\label{sec:extinction}

\begin{figure}[htp]

\includegraphics[width=\linewidth]{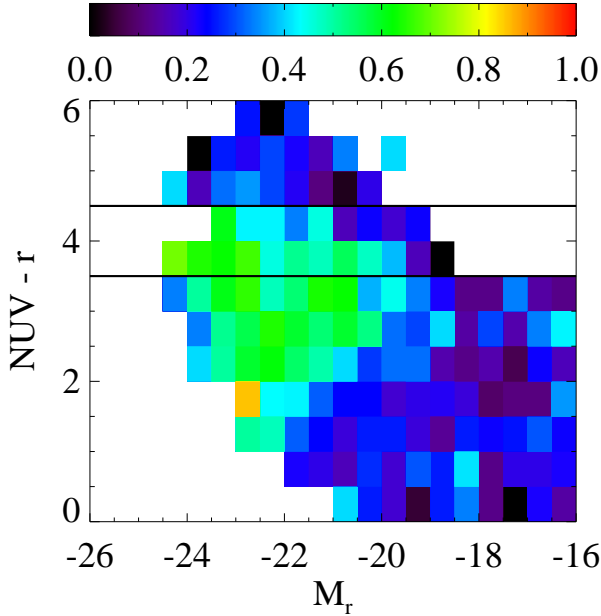}

\caption[Fraction of 24 $\mu$m-detected galaxies in the color-magnitude diagram]{Fraction of 24 $\mu$m detections as a function of color and magnitude. The thick horizontal lines indication the region of the color-magnitude diagram defined as the green valley in this work. Up to 65\% of the green valley galaxies in a given bin are probably dusty starbursts, and not galaxies in the process of quenching star formation.}\label{fig:24um}

\end{figure}

Extinction correction is of fundamental importance to identify quenching galaxies in the green valley. This is mainly due to contamination by heavily obscured star-forming galaxies. This is already an important problem at $z\sim 0$, but we expect it to be worse at higher redshifts, since the number density of LIRGs and ULIRGs is found to increase towards earlier times \citep{LeFloch2005,Magnelli2009}. To illustrate the issue, we show in Figure \ref{fig:24um} the fraction of 24 $\mu$m detected galaxies in the Groth Strip per two-dimensional bin in the color magnitude diagram \citep[for a description of the dataset, see][]{Davis2007}. We can see that most of the green valley galaxies on the bright end (up to $\sim 65$\% in some color-magnitude bins) are also 24 $\mu$m sources, indicative of dusty star forming galaxies, instead of the quenching objects for which we are searching.

\begin{figure}[htp]

\includegraphics[width=\linewidth]{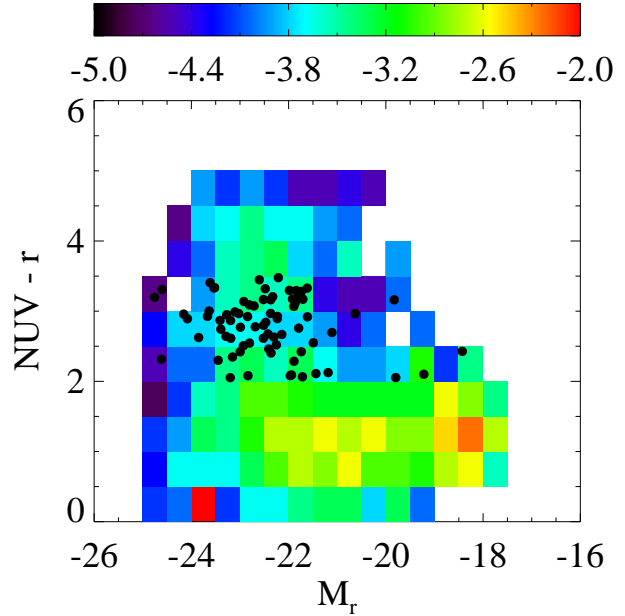}

\caption[Number density of galaxies in the extinction-corrected color-magnitude diagram]{Same as Figure \ref{fig:density}, after applying extinction correction of absolute magnitudes (see section \ref{sec:extinction}).}\label{fig:density_ext}

\end{figure}

In an attempt to decrease contamination from such galaxies, we used an independent SED fitting result for a number of galaxies in the Extended Groth Strip \citep{Salim2009}, where extinction is a free parameter. This sample of $\sim$6000 objects was then used to calculate the number densities in the color-magnitude diagram the same way as described in section \ref{sec:density}. The result can be found in Figure \ref{fig:density_ext}. The main difference when comparing this diagram with Figure \ref{fig:density} is a decrease in number density in the green valley and a clearer distinction between the blue and red sequences, as expected. We have also plotted the corresponding luminosity function for extinction-corrected green valley galaxies in Figure \ref{fig:lumfunction}. There is a decrease in $\Phi^*$, as expected; in addition, we notice a brighter value of $M^*$, since this represents the magnitude of galaxies corrected for internal extinction.

\section{Results}\label{sec:gvresults}

\begin{figure}[htp]

\includegraphics[width=\linewidth]{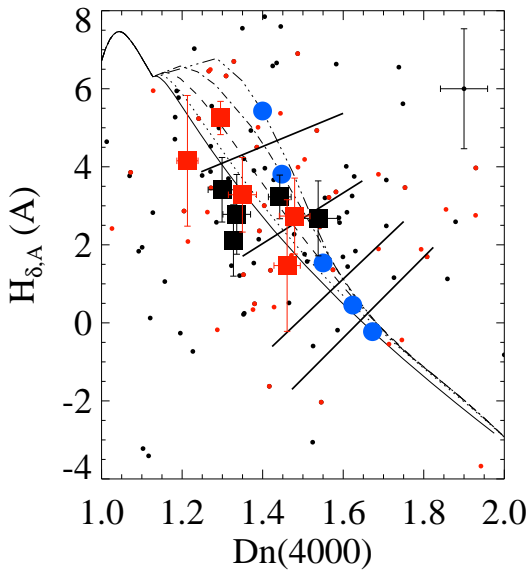}

\caption[D$_n(4000)$ and H$_{\delta,A}$ indices of green valley galaxies]{D$_n(4000)$ and H$_{\delta,A}$ indices of green valley galaxies with errorbars. The different lines on the plot indicate the expected values from models with varying $\gamma$ values, as in Figure \ref{fig:dn_hd_model}. Black points are all galaxies that fall in the green valley region, prior to making a correction for extinction in the selection. Red symbols indicate galaxies in the green valley after the extinction correction is applied. Large filled squares are the results for stacked spectra, color-coded as for individual galaxies (see section \ref{sec:stack}). Big blue circles indicate the moment at which models present ${\rm NUV}-r = 4.0$ colors, with diagonal straight lines representing equidistance from consecutive models. Also shown on the top right is the median error for each individual galaxy.}\label{fig:gv_dn_hd}

\end{figure}

We have measured $D_n(4000)$ and $H_{\delta,A}$ indices for all green valley galaxies observed. A number of galaxies presented unrealistic values -- outside the expected range of $1.0 < D_n(4000) < 2.0$ and $-4.0 < H_{\delta,A} < 8.0$. These objects present a number of issues, including signal-to-noise ratios $S/N<1$ per pixel, or they are too close to the edge of a mask and a proper extraction was not possible. Excluding these, we are left with 105 green valley galaxies. These objects are shown in Figure \ref{fig:gv_dn_hd}, along with the models presented in Figure \ref{fig:dn_hd_model}. The observed values agree well with the models, following the trend of lower $H_{\delta,A}$ values for higher $D_n(4000)$.

In selecting galaxies for the extinction-corrected green valley, our sample was reduced, due to smaller number densities and only a fraction (approximately 50\%) of galaxies in DEEP2 (and no galaxies in the remaining CFHTLS fields) having reliable SED extinction measurements. For this reason, we have expanded the color range to include $3.0 < ({\rm NUV}-r)_{\rm ext} < 4.5$ colors, in an attempt to improve our statistics. Extinction-corrected galaxies are highlighted in Figure \ref{fig:gv_dn_hd} as red symbols.

Figure \ref{fig:gv_dn_hd} also shows the D$_n(4000)$ and H$_{\delta,A}$ values for models of green valley galaxies defined at ${\rm NUV}-r = 4.0$, the median color value of our green valley sample. The straight diagonal lines indicate equidistant lines from two adjacent models, in which distance is defined by normalizing the H$_{\delta,A}$ index as H$_{\delta,An}=H_{\delta,A} / 12$ to reflect the expected dynamic range in that measurement.

\begin{figure}[htp]

\includegraphics[width=\linewidth]{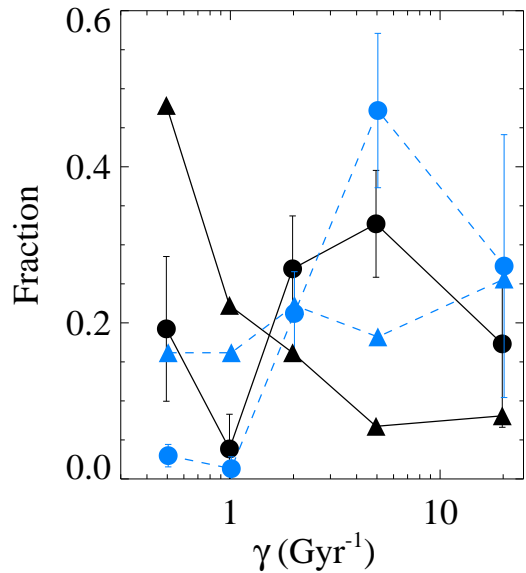}

\caption[Fraction of galaxies as a function of $\gamma$]{Fraction of green valley galaxies as a function of $\gamma$ bins, as shown in Figure \ref{fig:gv_dn_hd}. All values shown here are for extinction corrected samples; circles represent our sample, while the triangles are the values found at redshift $z\sim 0.1$ in Paper I. The blue dashed lines in both cases are weighted by $dy/dt$ to correct for the fact that galaxies that are quenching faster are less likely to be observed. We notice an evolution in the fraction of high-$\gamma$ values, in that at higher redshift the timescales for star formation quenching were shorter.}\label{fig:gamma_fraction}

\end{figure}

In Figure \ref{fig:gamma_fraction} we compare the quenching timescales we find with the obtained values at $z=0.1$ in Paper I, showing the fraction of galaxies in each $\gamma$ bin as defined by the dividing lines in Figure \ref{fig:gv_dn_hd}. Error bars represent fluctuations in each bin, when randomly shifting individual D$_n(4000)$ and H$_{\delta,A}$ measurements according to their corresponding errors; the error bar in each $\gamma$ bin is then the standard deviation in number counts after 1000 Monte Carlo simulations. Since $\gamma$ correlates with quenching speed, that means galaxies with higher $\gamma$ values will spend less time in the green valley, and are less likely to be observed. We show the fractions corrected for this (weighted by $dy/dt$) as dashed lines. We represent the data in this work as circles, and the values for low redshift as triangles. The main conclusion we draw from this exercise is that quenching timescales are shorter at higher redshift, since the amount of galaxies with higher $\gamma$ is increased. In quantitative terms, this represents a factor $2-3$ decrease in typical transitioning timescales.

We determined average $D_{n}(4000)$ and H$_{\delta,A}$ values for each magnitude bin by averaging over all galaxies in a given bin, weighted by the error in each case. This yields an average $\gamma$ in each bin, which in turn corresponds to a period of time required to cross over the color range covered by the green valley. We combine median galaxy masses for each bin, as determined from the K-band measurements by \citet{Bundy2007}, and number densities to calculate a mass flux for each given magnitude. We show these values in Table \ref{table:rho_dot}. We repeat the procedure in the case of the extinction-corrected green valley galaxies, and show the results in Table \ref{table:rho_dot_ext}.

\begin{table*}%{lcccccccc}
%\begin{deluxetable*}{lccccccc}
\begin{center}
%\tablewidth\linewidth
%\tabletypesize{\scriptsize}
%\tablecaption{Mass flux results}
\caption{Mass flux results}
\begin{tabular}{l c c c c c c c}\hline
%\tablehead{
%\colhead{$M_r$} & \colhead{log $\Phi$} & \colhead{$<\log M_*>$} & \colhead{n} & \colhead{$<D_n(4000)>$} & \colhead{$<H_{\delta,A}>$} & \colhead{$<\gamma>$} & \colhead{$\dot\rho$}\\
%  & \colhead{(Mpc$^{-3}$)} & \colhead{M$_\odot$} & & & & & \colhead{(10$^{-3}$ M$_\odot$ yr$^{-1}$ Mpc$^{-3}$)}\\
$M_r$ & log $\Phi$ & $<\log M_*>$ & n & $<D_n(4000)>$ & $<H_{\delta,A}>$ & $<\gamma>$ & $\dot\rho$\\
  & (Mpc$^{-3}$) & M$_\odot$ & & & & & (10$^{-3}$ M$_\odot$ yr$^{-1}$ Mpc$^{-3}$)\\
%  \label{table:rho_dot}
%  }
%\startdata
\hline
-24.25 & $-$5.32$\pm$0.59 & 11.3 & 1 & 1.54 & 2.43 & 3.05 & 2.0$\pm$2.7\\
-23.75 & $-$4.72$\pm$0.30 & 11.3 & 2 & 1.21 & $-$1.10 & 2.35 & 8.1$\pm$5.5\\
-23.25 & $-$4.08$\pm$0.14 & 11.2 & 5 & 1.24 & $-$0.41 & 2.60 & 30.0$\pm$9.8\\
-22.75 & $-$3.75$\pm$0.10 & 11.1 & 18 & 1.26 & 1.98 & 6.10 & 87.5$\pm$19.5\\
-22.25 & $-$3.51$\pm$0.07 & 11.0 & 23 & 1.32 & 2.61 & 6.95 & 99.8$\pm$17.1\\
-21.75 & $-$3.45$\pm$0.07 & 10.9 & 27 & 1.30 & 4.18 & 12.80 & 93.4$\pm$15.1\\
-21.25 & $-$3.41$\pm$0.08 & 10.7 & 15 & 1.32 & 4.43 & 13.39 & 65.9$\pm$11.7\\
-20.75 & $-$3.44$\pm$0.10 & 10.5 & 12 & 1.37 & 0.64 & 2.63 & 25.0$\pm$5.9\\
-20.25 & $-$3.59$\pm$0.15 & 10.2 & 3 & 1.33 & 4.18 & 12.32 & 17.7$\pm$6.3\\
-19.75 & $-$3.71$\pm$0.30 & 10.1 & 0 & N/A & N/A & N/A & N/A\\
-19.25 & $-$4.08$\pm$0.94 & 10.0 & 1 & 1.37 & 6.37 & 20.63 & 2.9$\pm$6.2\\
\hline
%\enddata
\label{table:rho_dot}
%\end{deluxetable*}
\end{tabular}
\end{center}
\end{table*}

\begin{table*}%{lcccccccc}
\begin{center}
%\tablewidth\linewidth
%\tabletypesize{\scriptsize}
\caption{Mass flux results (corrected for extinction)} 
\begin{tabular}{l c c c c c c c}\hline
%\tablehead{
$M_r$ & log $\Phi$ & $<\log M_*>$ & n & $<D_n(4000)>$ & $<H_{\delta,A}>$ & $<\gamma>$ & $\dot\rho$\\
  & (Mpc$^{-3}$) & M$_\odot$ & & & & & (10$^{-3}$ M$_\odot$ yr$^{-1}$ Mpc$^{-3}$)\\
%  }
%\startdata
\hline
-24.75 & $-$3.91$\pm$0.57 & 11.1 & 1 & 1.71 & 2.37 & 1.68 & 26.8$\pm$35.4\\
-24.25 & $-$3.83$\pm$0.37 & 11.1 & 0 & N/A & N/A & N/A & N/A\\
-23.75 & $-$3.93$\pm$0.34 & 11.1 & 3 & 1.45 & 1.73 & 2.98 & 32.3$\pm$25.3\\
-23.25 & $-$3.92$\pm$0.35 & 11.1 & 12 & 1.18 & $-$1.11 & 2.47 & 31.1$\pm$24.9\\
-22.75 & $-$3.92$\pm$0.49 & 10.9 & 9 & 1.33 & 2.43 & 6.17 & 37.3$\pm$41.9\\
-22.25 & $-$4.28$\pm$0.53 & 10.8 & 10 & 1.41 & 1.21 & 2.80 & 7.4$\pm$8.9\\
-21.75 & $-$4.09$\pm$1.33 & 10.7 & 9 & 1.35 & 2.35 & 5.63 & 13.3$\pm$40.8\\
-21.25 & $-$4.11$\pm$1.02 & 10.5 & 5 & 1.21 & 4.64 & 15.72 & 9.1$\pm$21.3\\
\hline
%\enddata
\label{table:rho_dot_ext}
\end{tabular}
\end{center}
%\tablenotetext{a}{in mas}
%\tablenotetext{b}{UV half-light radius from HST data}
\end{table*}

\begin{figure}[htp]

\includegraphics[width=\linewidth]{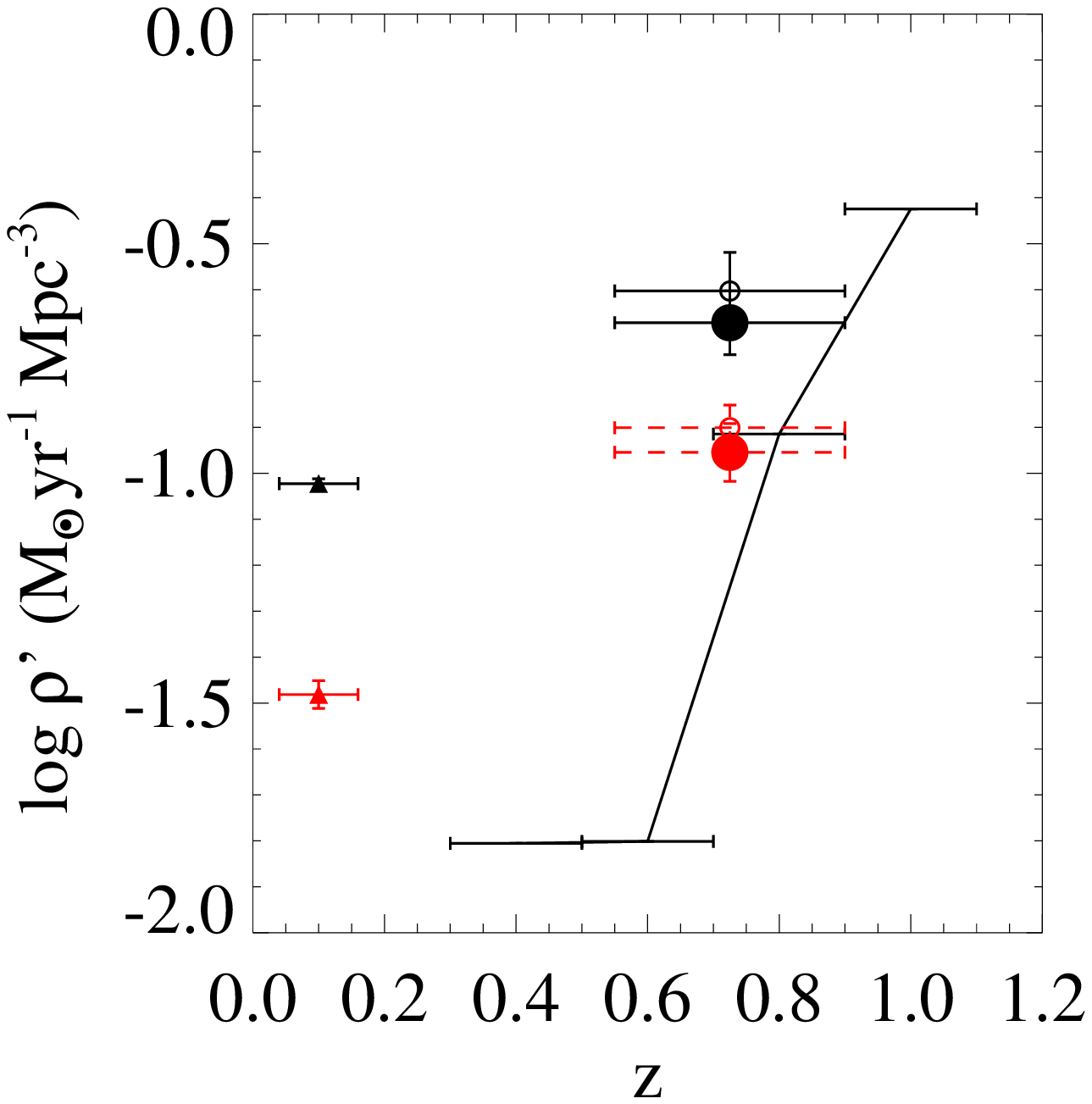}

\caption[Evolution of the mass flux density through the green valley]{Mass flux density in the green valley. The circles represent our data; solid error bars indicate the value calculated without extinction correction, while the red symbols with dashed error bars include extinction correction. Empty circles indicate values measured from stacked spectra (see section \ref{sec:stack}). Values at $z=0.1$ calculated in Paper I are shown as triangles, and are determined with (red) and without (black) extinction correction. The solid line shows the density growth rate of the red sequence, as determined from \citet{Faber2007}. Our data points show a clear increase in the mass flux density across the green valley towards earlier times, in agreement with estimates from the growth of the red sequence.}\label{fig:mass_flux}

\end{figure}

Finally, the total mass flux density is the sum of the mass flux through all magnitudes. We show the results in Figure \ref{fig:mass_flux}. The upper circle represents the value before extinction correction ($\log\dot\rho=-0.36\pm0.08$ M$_\odot$ yr$^{-1}$ Mpc$^{-3}$), while the bottom value represents the mass flux while accounting for extinction correction ($\log\dot\rho_{\rm ext}=-0.80\pm0.51$ M$_\odot$ yr$^{-1}$ Mpc$^{-3}$). The horizontal error bars represent the redshift range covered by our sample, while vertical error bars include errors in number density and transition timescales as described in previous sections.

The first thing to notice is the difference (about a factor of 3) between both measurements. This can be attributed to a combination of factors: on one hand, measured number densities are smaller when correction for extinction is applied (section \ref{sec:extinction}); on the other hand, the contamination of star-forming galaxies also biases our sample to higher $<D_n(4000)>$, $<H_{\delta,A}>$ values, which in turn represent higher $\gamma$'s and shorter evolution timescales.

We compare our values to those found in Paper I at $z=0.1$; the top triangle in Figure \ref{fig:mass_flux} represents the flux without taking into account dust extinction, and the bottom one is the extinction-corrected measurement. The evolution with redshift from $z=0.1$ to $0.8$ is evident, with mass flux values at intermediate redshifts being 3 to 5 times higher than those found in the low-redshift universe. This reflects the significant change occurring in galaxy evolution over cosmic time.

In Figure \ref{fig:mass_flux} we also show the density growth of the red sequence, in units of M$_\odot$ yr$^{-1}$ Mpc$^{-3}$. This has been calculated as follows: the B-band luminosity density at the local universe has been determined as $j_B=10^{7.7}$ L$_\odot$ Mpc$^{-3}$ \citep{Bell2003,Madgwick2002}. The mass-to-light ratio of red sequence galaxies in the B band, in turn, is estimated at $({\rm M}/L_B) = 6$ M$_\odot/L_\odot$, using a constant, average value for the redshift range covered \citep{Gebhardt2003, Faber2007}. Combined with the evolution in the number density of galaxies in the red sequence \citep{Faber2007}, we can then calculate the mass flux at each redshift interval as

\begin{equation}
\dot\rho=j_B\left(\frac{\rm M}{L_B}\right)\left(\frac{10^{\Delta\log\Phi_0}-1}{10^{\Delta\log\Phi_0}}\right)\Delta t(\Delta z),
\end{equation}
where $\Delta\log\Phi_0$ is the logarithmic change in the normalization of the luminosity function at each redshift, and $\Delta t(\Delta z)$ is the change in the age of the universe at each redshift interval.

We find that the mass growth rate was indeed higher by a factor of $\sim 5$ in the earlier universe, pointing to another manifestation of downsizing in galaxy evolution. Furthermore, our results agree within errors with the estimated growth of the red sequence at $z\sim 0.8$, reinforcing the idea that the mass flux through the green valley occurred more rapidly in the past.

\subsection{Stacked spectra}\label{sec:stack}

\begin{figure}

\includegraphics[width=\linewidth]{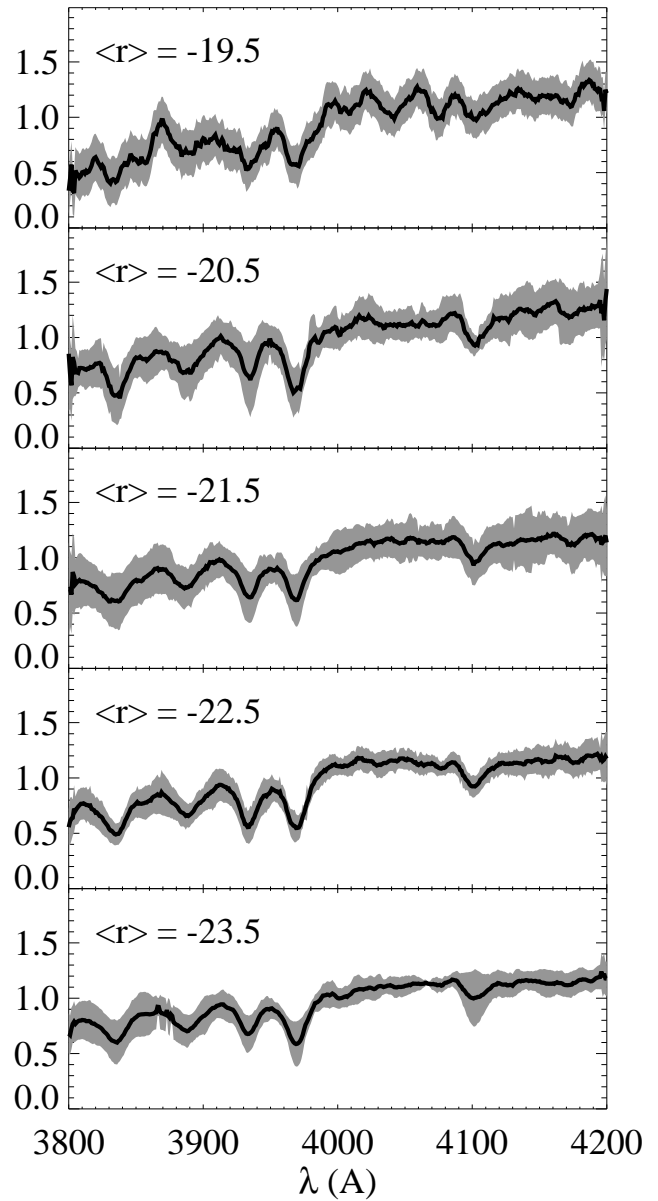}

\caption[Evolution of the mass flux density through the green valley]{Stacked spectra for each magnitude bin. Grey shades indicate the standard deviation in each pixel. In all cases spectral features are clearly distinguishable, in particular the measured absorption in H$_\delta$ at 4100 \AA.}\label{fig:coadded_spectra}

\end{figure}

In order to optimize the S/N of the spectral features our results are based on, we have produced coadded spectra within multiple absolute magnitude bins and repeated the same analysis we applied to our individual spectra. The magnitude bins, each 1mag in width, were defined to be representative of the sample: M$_{-24,-23}$, M$_{-23,-22}$, M$_{-22,-21}$, M$_{-21,-20}$,M$_{-20,-19}$. We combine all individual spectra in our sample within each absolute magnitude bin performing a {\it meanclip} averaging for each wavelength pixel between 3800 and 4200 {\AA}. The resulting {\it stacked spectra} are shown in Figure \ref{fig:coadded_spectra}, with the shaded region indicating the standard deviation in each pixel. With a markedly higher S/N, the spectral features in our coadded spectra are clearly distinguishable. To ensure comparison on equal footing with results based on individual spectra, we also produced stacked specra for absolute magnitudes after applying a correction for extinction: M$_{-25,-24}$, M$_{-24,-23}$, M$_{-23,-22}$, M$_{-22,-21}$,M$_{-21,-20}$.

Based on the stacked spectra we measure mass flux within each magnitude bin repeating the same procedure applied to the individual spectra. The measured values of D$_{n}(4000)$ and H$_{\delta,A}$ are shown in Figure \ref{fig:gv_dn_hd} as filled squares. With the improvement in S/N, the results based on stacked spectra show a substantial increase in agreement with the models, strengthening our earlier results. We confirm that green valley galaxies show lower values for D$_n(4000)$ and stronger absorption in H$_\delta$, indicating a stellar population younger than that of their low-redshift counterparts, suggesting faster migration from the blue cloud to the red clump at higher redshifts.

The sum of the mass flux through the green valley based on the stacked spectra analysis before and after extinction is shown in Figure \ref{fig:mass_flux} as empty circles, following the same color code as before. The resulting values agree with those measured for the average of individual galaxies within 1-$\sigma$, confirming our results: the mass flux density across the green valley at redshift $z\sim 0.8$ is greater than found at low-redshift ($z\sim 0.1$) by a factor of $\sim 5$.

The sum of the mass flux through the green valley measured this way is shown in Figure \ref{fig:mass_flux} as hollow circles, color-coded as before. The values agree with those measured for the average of individual galaxies within 1-$\sigma$, confirming our results: the mass flux density across the green valley at redshift $z\sim 0.8$ is greater than found at low-redshift ($z\sim 0.1$) by a factor of $\sim 5$.

\section{Discussion}\label{sec:gvdiscussion}

\subsection{The build-up of the red sequence}

In \citet{Faber2007}, the authors discuss how the red sequence at $z=0$ is assembled, from the brightest, most massive objects down to the smallest and faintest. From DEEP2 and COMBO17 data out to $z\sim 1$, the authors are able to measure the evolution of the luminosity function and luminosity density -- for all galaxies and for the blue and red sequences individually -- to discuss how galaxies might evolve along the color-magnitude diagram. The authors argue that the red sequence is formed through a combination of star formation quenching in star forming galaxies and dry mergers, which move galaxies along the red sequence towards the brighter end.

How are the brightest red galaxies formed? One might think that the simple quenching of individual blue galaxies, be it by mergers, AGN activity or any other mechanism, would be sufficient to produce the red sequence. However, the most massive blue galaxies in the local universe are not as massive as the most massive red ones. Even major mergers, which would produce a galaxy up to twice as massive as each individual object, are not enough to explain the observed mass function. In fact, in the morphological study by Mendez et al. 2011 of $\sim$300 galaxies in the optically-defined green valley at $0.4 < z < 1.2$, they find that most green galaxies cannot be classified as mergers and that, the merger fraction in the green valley is in fact lower than in the blue sequence.

\begin{figure}[htp]

\includegraphics[width=\linewidth]{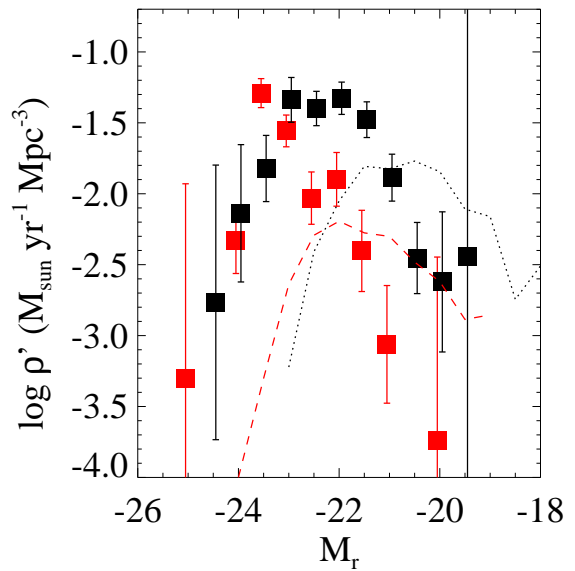}

\caption[Mass flux density as a function of $r$ magnitude]{Mass flux density as a function of $r$ magnitude. The flux at intermediate redshifts is shown as solid symbols; red triangles for the extinction-corrected sample, black squares, no extinction correction. The lines represents the values found for the local universe in Paper I; the red dashed line indicates the extinction corrected sample at $z=0.1$. In both cases, the peak at high redshift is shifted towards brighter magnitudes by $\gtrsim 1$ mag.}\label{fig:mass_flux_vs_mag}

\end{figure}

Our data offers an interesting insight into the problem of red galaxy formation. Comparison with the mass-flux at low-redshift (Paper I) shows that the mass flux occurred at brighter magnitudes at high redshift (Figure \ref{fig:mass_flux_vs_mag}), indicating the build-up of the most massive end of the red sequence at earlier times, which is in qualitative agreement with the evolution of the luminosity function. Our sample reliably covers magnitudes down to $r\sim -20$ (not accounting for extinction correction) and $r\sim -22$ (accounting for extinction correction), allowing us to verify the decrease in number density and mass flux towards fainter objects (see Figure \ref{fig:lumfunction}). This represents a {\it downsizing} of the green valley evolution, with the red sequence forming ``from the top down'': in the past, more massive star-forming galaxies were being formed and subsequently quenched, forming the more massive red sequence galaxies. At later times, star formation shifts to less massive galaxies; these are then quenched as well, and the fainter end of the red sequence is created.

It has been argued before that the massive galaxies in the red sequence form in earlier times. As an example, \citet{Borch2006} have measured the stellar mass function of the red sequence up to redshift $z\sim 1.0$ and have noticed that the number density of less massive galaxies has grown more than the number density of massive galaxies since that epoch. The evolution we observe in the luminosity function supports this idea, with a change in the faind-end slope that indicates a deficit of faint, less massive galaxies when compared to the same population at lower redshift. The analysis of objects in the green valley, on the other hand, describes this scenario for the first time based on observations of the mass flux of intermediate galaxies alone. 

However, this is not the complete picture. \citep{Faber2007} call upon dry mergers to explain observed properties of the brightest red galaxies. These mergers correspond to the interaction between two or more red sequence galaxies, with little gas involved. The absence of copious gas diminishes the subsequent burst in star formation that is otherwise expected in wet mergers, between gas-rich galaxies. These dry mergers hence result in an increase in stellar mass, with no coupled burst in activity. It would be interesting to compare the mass flux density at the very massive end of the color-magnitude diagram to infer whether that is sufficient to create all massive galaxies or if dry mergers are a necessity, but this is currently an impossible task due to large uncertainties and small number statistics (especially at these limiting magnitudes).

It could be argued that changes in galaxy properties related to an increase in the mass-to-light ratio could account for the dimming of red sequence galaxies, in such a way that galaxies of a given stellar mass would appear fainter at recent epochs. While it is true that the mass-to-light ratio increases as a function of time \citep[e.g.][]{Bell2004}, the change in the $r$-band is not large enough to account for the evolution of the luminosity function by itself, especially for passive galaxies. We have measured the mass-to-light ratio of galaxies of a {\rm constant color} $4.5<NUV-r<5.0$ as a function of redshift, to evaluate any possible evolution within the color range covered by the luminosity functions, based on the K-band measurements from \citet{Bundy2006}. The average value between $0.1<z<0.3$ is $(M/L_r)=1.9\pm 0.6$, while at higher redshifts, $0.8<z<1.0$, $(M/L_r)=1.8\pm 0.6$. We can conclude, then, that the increase in mass-to-light ratios of red galaxies is not high enough to account for the evolution in the luminosity function, and galaxies in that color range are indeed more massive at higher redshift.

The mechanism through which quenching occurs is as yet unclear. As we have seen above, merger activity may not be as relevant; AGN activity is also called upon, and it is indeed found that green valley galaxies show an increase in AGN fraction -- although actual correlation with star formation quenching has not been unequivocally shown \citep[Paper I;][]{Nandra2007,Schawinski2009}. In a future paper, we plan to investigate this correlation by comparing our results for quenching timescales with different tracers of AGN activity, such as X-ray luminosities and optical emission line ratios.

\subsection{The star formation history of green valley galaxies}\label{sec:sfh}

It is important to consider, first, that most of the stellar mass in red sequence galaxies is not formed while the galaxy is in the red sequence \citep{Salim2005}. This lends support to our assumption (1), which states that galaxies are only moving redward in the color-magnitude diagram. However, this is not strictly true; we do know a fraction of elliptical galaxies, even in the local universe, show signs of recent starburst events \citep{Thilker2010}. If we take into account that a fraction of galaxies in the green valley are turning {\it blue}, instead, then the inferred mass flux would be {\it smaller}.

At the same time, in Paper I we have argued how our choice of star formation histories influences our results. In general, assuming shorter periods of constant star formation or different models for the time evolution decreases the final calculated value for $\dot\rho$. Therefore, we conservatively assume our measured mass flux density as an {\it upper value} for the actual flux in transition galaxies from the blue to the red sequence. Still, a decrease by 0.5 dex brings us to closer agreement with the inferred growth of the red sequence. At the same time, the difference in final results combined with the observed change in the luminosity function and the distribution of $\gamma$ values makes it safe to assume the evolution in mass flux density from $z\sim 0.8$ to $z\sim 0.1$ is real.

Finally, one can argue the star formation history assumed in this work is simplistic. It has been even argued that star forming galaxies follow an inverted $\tau$ model, with an exponentially {\it increasing} star formation history \citep{Maraston2010}. In any case, a more realistic approach to modeling the star formation histories of galaxies could yield different results (likely smaller fluxes, as argued above). Thus, we are currently in the process of producing models that do not rely on {\it ad hoc} star formation histories, but instead are drawn from cosmological N-body simulations and semi-analytic models, generating more physically motivated star formation histories. We will compare our spectra with these more refined models in a future paper (Martin et al., in prep.).

\section{Summary}\label{sec:gvsummary}

We have shown results of a spectroscopic survey of over 100 galaxies in the green valley, i.e. transition objects with intermediate colors between the blue and the red sequences that are believed to be currently quenching their star formation. The data represent the deepest spectra of green valley galaxies ever obtained at intermediate redshifts ($z\sim 0.8$).

By using measurements of spectral indices - namely the break at 4000 {\AA} and the equivalent width of the H$_\delta$ absorption line, we are able to infer the star formation histories of these objects, following the method first presented in \citet{Martin2007}. If one assumes an exponential decline in star formation rates, it is then possible to measure timescales involved in transitioning from the population of star-forming galaxies into the red sequence. Combined with independent measurements for the number densities in the green valley and typical galaxy masses, we are able to measure the mass flux density between $0.55<z<0.9$.

Our measurements have shown that the mass flux density at these redshifts is higher by a factor of $\sim$5 than in the low-redshift universe ($z\sim 0.1$). This can be attributed to two factors: first, the evolution of the luminosity function over the course of the last 6 Gyr, which means at $z\sim 0.8$ the flux is dominated by the brighter and more massive objects. Second, quenching timescales are shorter at higher redshift, and galaxies transition more rapidly from the blue to the red sequence. In addition, we also show that our results are in good agreement with estimates for the build-up of mass in the red sequence since $z=1$. We argue for a scenario in which the red-sequence is built ``from the top-down'', meaning that the most massive objects were quenched at earlier times, shifting towards the evolution of less massive galaxies today.

Since we have calculated the quenching timescales of individual objects, the next logical step is to compare our results with other observable properties in the green valley galaxies, in particular merger signatures and AGN activity. We expect this analysis will help clarify the role of each of these processes in the total quenching of star formation, both at low and at high redshifts.

\vskip .5in

We thank the anonymous referee for comments that have significantly improved this paper, with special attention to the luminosity functions and coadded spectra. TSG would like to thank Samir Salim for useful comments and suggestions. Some of the data presented herein were obtained at the W.M. Keck Observatory, which is operated as a scientific partnership among the California Institute of Technology, the University of California and the National Aeronautics and Space Administration. The Observatory was made possible by the generous financial support of the W.M. Keck Foundation. The authors wish to recognize and acknowledge the very significant cultural role and reverence that the summit of Mauna Kea has always had within the indigenous Hawaiian community.  We are most fortunate to have the opportunity to conduct observations from this mountain. This study makes use of data from AEGIS, a multiwavelength sky survey conducted with the Chandra, GALEX, Hubble, Keck, CFHT, MMT, Subaru, Palomar, Spitzer, VLA, and other telescopes and supported in part by the NSF, NASA, and the STFC. The analysis pipeline used to reduce the DEIMOS data was developed at UC Berkeley with support from NSF grant AST-0071048. Based on observations obtained with MegaPrime/MegaCam, a joint project of CFHT and CEA/DAPNIA, at the Canada-France-Hawaii Telescope (CFHT) which is operated by the National Research Council (NRC) of Canada, the Institut National des Science de l'Univers of the Centre National de la Recherche Scientifique (CNRS) of France, and the University of Hawaii. This work is based in part on data products produced at TERAPIX and the Canadian Astronomy Data Centre as part of the Canada-France-Hawaii Telescope Legacy Survey, a collaborative project of NRC and CNRS.

\clearpage
\setlength\headsep{0.8in}


\begin{thebibliography}{65}
\expandafter\ifx\csname natexlab\endcsname\relax\def\natexlab#1{#1}\fi

\bibitem[{Aird {et~al.}(2012)Aird, Coil, Moustakas, Blanton, Burles, Cool,
  Eisenstein, Smith, Wong, \& Zhu}]{Aird2012}
Aird, J., Coil, A.~L., Moustakas, J., {et~al.} 2012, \apj, 746, 90

\bibitem[{Baldry {et~al.}(2004)Baldry, Glazebrook, Brinkmann, Ivezi\'{c},
  Lupton, Nichol, \& Szalay}]{Baldry2004}
Baldry, I.~K., Glazebrook, K., Brinkmann, J., {et~al.} 2004, \apj, 600, 681

\bibitem[{Balogh {et~al.}(1999)Balogh, Morris, Yee, Carlberg, \&
  Ellingson}]{Balogh1999}
Balogh, M.~L., Morris, S.~L., Yee, H. K.~C., Carlberg, R.~G., \& Ellingson, E.
  1999, \apj, 527, 54

\bibitem[{Balogh {et~al.}(1998)Balogh, Schade, Morris, Yee, Carlberg, \&
  Ellingson}]{Balogh1998}
Balogh, M.~L., Schade, D., Morris, S.~L., {et~al.} 1998, \apj, 504, L75

\bibitem[{Bell {et~al.}(2003)Bell, McIntosh, Katz, \& Weinberg}]{Bell2003}
Bell, E.~F., McIntosh, D.~H., Katz, N., \& Weinberg, M.~D. 2003, \apjs, 149,
  289

\bibitem[{Bell {et~al.}(2004)Bell, Wolf, Meisenheimer, Rix, Borch, Dye,
  Kleinheinrich, Wisotzki, \& McIntosh}]{Bell2004}
Bell, E.~F., Wolf, C., Meisenheimer, K., {et~al.} 2004, \apj, 608, 752

\bibitem[{Blanton {et~al.}(2005)Blanton, Eisenstein, Hogg, Schlegel, \&
  Brinkmann}]{Blanton2005}
Blanton, M.~R., Eisenstein, D., Hogg, D.~W., Schlegel, D.~J., \& Brinkmann, J.
  2005, \apj, 629, 143

\bibitem[{Blanton \& Roweis(2007)}]{Blanton2007}
Blanton, M.~R., \& Roweis, S. 2007, \aj, 133, 734

\bibitem[{Borch {et~al.}(2006)Borch, Meisenheimer, Bell, Rix, Wolf, Dye,
  Kleinheinrich, Kovacs, \& Wisotzki}]{Borch2006}
Borch, A., Meisenheimer, K., Bell, E.~F., {et~al.} 2006, A\&A, 453, 869

\bibitem[{Bruzual(1983)}]{BruzualA.1983}
Bruzual, G. 1983, \apj, 273, 105

\bibitem[{Bruzual \& Charlot(2003)}]{Bruzual2003}
Bruzual, G., \& Charlot, S. 2003, \mnras, 344, 1000

\bibitem[{Bundy {et~al.}(2007)Bundy, Treu, \& Ellis}]{Bundy2007}
Bundy, K., Treu, T., \& Ellis, R.~S. 2007, \apj, 665, L5

\bibitem[{Bundy {et~al.}(2006)Bundy, Ellis, Conselice, Taylor, Cooper, Willmer,
  Weiner, Coil, Noeske, \& Eisenhardt}]{Bundy2006}
Bundy, K., Ellis, R.~S., Conselice, C.~J., {et~al.} 2006, \apj, 651, 120

\bibitem[{Catinella {et~al.}(2010)Catinella, Schiminovich, Kauffmann, Fabello,
  Wang, Hummels, Lemonias, Moran, Wu, Giovanelli, Haynes, Heckman, Basu-Zych,
  Blanton, Brinchmann, Budav\'{a}ri, Gon\c{c}alves, Johnson, Kennicutt, Madore,
  Martin, Rich, Tacconi, Thilker, Wild, \& Wyder}]{Catinella2010}
Catinella, B., Schiminovich, D., Kauffmann, G., {et~al.} 2010, \mnras, 403, 683

\bibitem[{Chabrier(2003)}]{Chabrier2003}
Chabrier, G. 2003, PASP, 115, 763

\bibitem[{Coil {et~al.}(2008)Coil, Newman, Croton, Cooper, Davis, Faber, Gerke,
  Koo, Padmanabhan, Wechsler, \& Weiner}]{Coil2008}
Coil, A.~L., Newman, J.~A., Croton, D., {et~al.} 2008, \apj, 672, 153

\bibitem[{Cooper {et~al.}(2006)Cooper, Newman, Croton, Weiner, Willmer, Gerke,
  Madgwick, Faber, Davis, Coil, Finkbeiner, Guhathakurta, \& Koo}]{Cooper2006}
Cooper, M.~C., Newman, J.~A., Croton, D.~J., {et~al.} 2006, \mnras, 370, 198

\bibitem[{Davis(2003)}]{Davis2003}
Davis, M. 2003, {Science Objectives and Early Results of the DEEP2 Redshift
  Survey}, Vol. 4834 (SPIE), 161--172

\bibitem[{Davis \& Geller(1976)}]{Davis1976}
Davis, M., \& Geller, M.~J. 1976, \apj, 208, 13

\bibitem[{Davis {et~al.}(2007)Davis, Guhathakurta, Konidaris, Newman, Ashby,
  Biggs, Barmby, Bundy, Chapman, Coil, Conselice, Cooper, Croton, Eisenhardt,
  Ellis, Faber, Fang, Fazio, Georgakakis, Gerke, Goss, Gwyn, Harker, Hopkins,
  Huang, Ivison, Kassin, Kirby, Koekemoer, Koo, Laird, {Le Floc'h}, Lin, Lotz,
  Marshall, Martin, Metevier, Moustakas, Nandra, Noeske, Papovich, Phillips,
  Rich, Rieke, Rigopoulou, Salim, Schiminovich, Simard, Smail, Small, Weiner,
  Willmer, Willner, Wilson, Wright, \& Yan}]{Davis2007}
Davis, M., Guhathakurta, P., Konidaris, N.~P., {et~al.} 2007, \apj, 660, L1

\bibitem[{{Di Matteo} {et~al.}(2005){Di Matteo}, Springel, \&
  Hernquist}]{DiMatteo2005}
{Di Matteo}, T., Springel, V., \& Hernquist, L. 2005, Nature, 433, 604

\bibitem[{Dressler(1980)}]{Dressler1980}
Dressler, A. 1980, \apj, 236, 351

\bibitem[{Faber(2003)}]{Faber2003}
Faber, S.~M. 2003, in Proceedings of SPIE, Vol. 4841 (SPIE), 1657--1669

\bibitem[{Faber {et~al.}(2007)Faber, Willmer, Wolf, Koo, Weiner, Newman, Im,
  Coil, Conroy, Cooper, Davis, Finkbeiner, Gerke, Gebhardt, Groth,
  Guhathakurta, Harker, Kaiser, Kassin, Kleinheinrich, Konidaris, Kron, Lin,
  Luppino, Madgwick, Meisenheimer, Noeske, Phillips, Sarajedini, Schiavon,
  Simard, Szalay, Vogt, \& Yan}]{Faber2007}
Faber, S.~M., Willmer, C. N.~A., Wolf, C., {et~al.} 2007, \apj, 665, 265

\bibitem[{Fortson {et~al.}(2011)Fortson, Masters, Nichol, Borne, Edmondson,
  Lintott, Raddick, Schawinski, \& Wallin}]{FortsonLucy2011}
Fortson, L., Masters, K., Nichol, R., {et~al.} 2012, in Advances in Machine Learning and Data Mining for Astronomy, ed. M. J. Way et al. (Chapman and Hall/CRC), 213

\bibitem[{Gebhardt {et~al.}(2003)Gebhardt, Faber, Koo, Im, Simard, Illingworth,
  Phillips, Sarajedini, Vogt, Weiner, \& Willmer}]{Gebhardt2003}
Gebhardt, K., Faber, S.~M., Koo, D.~C., {et~al.} 2003, \apj, 597, 239

\bibitem[{Giovanelli {et~al.}(1986)Giovanelli, Haynes, \&
  Chincarini}]{Giovanelli1986}
Giovanelli, R., Haynes, M.~P., \& Chincarini, G.~L. 1986, \apj, 300, 77

\bibitem[{Hubble(1926)}]{Hubble1926}
Hubble, E.~P. 1926, \apj, 64, 321

\bibitem[{Hubble(1927)}]{HubbleE.P.1927}
---. 1927, Obs, 50, 276

\bibitem[{Im {et~al.}(2002)Im, Simard, Faber, Koo, Gebhardt, Willmer, Phillips,
  Illingworth, Vogt, \& Sarajedini}]{Im2002}
Im, M., Simard, L., Faber, S.~M., {et~al.} 2002, \apj, 571, 136

\bibitem[{Kauffmann {et~al.}(2003)Kauffmann, Heckman, {Simon White}, Charlot,
  Tremonti, Brinchmann, Bruzual, Peng, Seibert, Bernardi, Blanton, Brinkmann,
  Castander, Cs\'{a}bai, Fukugita, Ivezic, Munn, Nichol, Padmanabhan, Thakar,
  Weinberg, \& York}]{Kauffmann2003}
Kauffmann, G., Heckman, T.~M., {Simon White}, D.~M., {et~al.} 2003, \mnras,
  341, 33

\bibitem[{Kaviraj {et~al.}(2008)Kaviraj, Khochfar, Schawinski, Yi, Gawiser,
  Silk, Virani, Cardamone, van Dokkum, \& Urry}]{Kaviraj2008}
Kaviraj, S., Khochfar, S., Schawinski, K., {et~al.} 2008, \mnras, 388, 67

\bibitem[{Koo {et~al.}(1996)Koo, Vogt, Phillips, Guzman, Wu, Faber, Gronwall,
  Forbes, Illingworth, Groth, Davis, Kron, \& Szalay}]{Koo1996}
Koo, D.~C., Vogt, N.~P., Phillips, A.~C., {et~al.} 1996, \apj, 469, 535

\bibitem[{Koyama {et~al.}(2011)Koyama, Kodama, Nakata, Shimasaku, \&
  Okamura}]{Koyama2011}
Koyama, Y., Kodama, T., Nakata, F., Shimasaku, K., \& Okamura, S. 2011, \apj, 734, 66

\bibitem[{{Le Floc’h} {et~al.}(2005){Le Floc’h}, Papovich, Dole, Bell,
  Lagache, Rieke, Egami, Perez‐Gonzalez, Alonso‐Herrero, Rieke, Blaylock,
  Engelbracht, Gordon, Hines, Misselt, Morrison, \& Mould}]{LeFloch2005}
{Le Floc’h}, E., Papovich, C., Dole, H., {et~al.} 2005, \apj, 632, 169

\bibitem[{L\'{o}pez-Sanjuan {et~al.}(2010)L\'{o}pez-Sanjuan, F\`{e}vre,
  de~Ravel, Cucciati, Ilbert, Tresse, Bardelli, Bolzonella, Contini, Garilli,
  Guzzo, Maccagni, McCraken, Mellier, Pollo, Vergani, \&
  Zucca}]{Lopez-Sanjuan2010}
L\'{o}pez-Sanjuan, C., F\`{e}vre, O.~L., de~Ravel, L., {et~al.} 2011, \aap, 530, A20

\bibitem[{Lotz {et~al.}(2008)Lotz, Davis, Faber, Guhathakurta, Gwyn, Huang,
  Koo, {Le Floc’h}, Lin, Newman, Noeske, Papovich, Willmer, Coil, Conselice,
  Cooper, Hopkins, Metevier, Primack, Rieke, \& Weiner}]{Lotz2008}
Lotz, J.~M., Davis, M., Faber, S.~M., {et~al.} 2008, \apj, 672, 177

\bibitem[{Madgwick {et~al.}(2002)Madgwick, Lahav, Baldry, Baugh,
  Bland-Hawthorn, Bridges, Cannon, Cole, Colless, Collins, Couch, Dalton, {De
  Propris}, Driver, Efstathiou, Ellis, Frenk, Glazebrook, Jackson, Lewis,
  Lumsden, Maddox, Norberg, Peacock, Peterson, Sutherland, \&
  Taylor}]{Madgwick2002}
Madgwick, D.~S., Lahav, O., Baldry, I.~K., {et~al.} 2002, \mnras, 333, 133

\bibitem[{Magnelli {et~al.}(2009)Magnelli, Elbaz, Chary, Dickinson, {Le
  Borgne}, Frayer, \& Willmer}]{Magnelli2009}
Magnelli, B., Elbaz, D., Chary, R.~R., {et~al.} 2009, A\&A, 496, 57

\bibitem[{Maraston {et~al.}(2010)Maraston, Pforr, Renzini, Daddi, Dickinson,
  Cimatti, \& Tonini}]{Maraston2010}
Maraston, C., Pforr, J., Renzini, A., {et~al.} 2010, \mnras, 407, 830

\bibitem[{Martin {et~al.}(2005)Martin, Fanson, Schiminovich, Morrissey,
  Friedman, Barlow, Conrow, Grange, Jelinsky, Milliard, Siegmund, Bianchi,
  Byun, Donas, Forster, Heckman, Lee, Madore, Malina, Neff, Rich, Small,
  Surber, Szalay, Welsh, \& Wyder}]{Martin2005}
Martin, D.~C., Fanson, J., Schiminovich, D., {et~al.} 2005, \apj, 619, L1

\bibitem[{Martin {et~al.}(2007)Martin, Wyder, Schiminovich, Barlow, Forster,
  Friedman, Morrissey, Neff, Seibert, Small, Welsh, Bianchi, Donas, Heckman,
  Lee, Madore, Milliard, Rich, Szalay, \& Yi}]{Martin2007}
Martin, D.~C., Wyder, T.~K., Schiminovich, D., {et~al.} 2007, \apjs, 173, 342

\bibitem[{Menci {et~al.}(2005)Menci, Fontana, Giallongo, \&
  Salimbeni}]{Menci2005}
Menci, N., Fontana, A., Giallongo, E., \& Salimbeni, S. 2005, \apj, 632, 49

\bibitem[{Mendez {et~al.}(2011)Mendez, Coil, Lotz, Salim, Moustakas, \&
  Simard}]{Mendez2011}
Mendez, A.~J., Coil, A.~L., Lotz, J., {et~al.} 2011, \apj, 736, 110

\bibitem[{Morganti {et~al.}(2006)Morganti, {De Zeeuw}, Oosterloo, McDermid,
  Krajnovi\'{c}, Cappellari, Kenn, Weijmans, \& Sarzi}]{Morganti2006}
Morganti, R., {De Zeeuw}, P.~T., Oosterloo, T.~A., {et~al.} 2006, \mnras, 371,
  157

\bibitem[{Nandra {et~al.}(2007)Nandra, Georgakakis, Willmer, Cooper, Croton,
  Davis, Faber, Koo, Laird, \& Newman}]{Nandra2007}
Nandra, K., Georgakakis, A., Willmer, C. N.~A., {et~al.} 2007, \apj, 660, L11

\bibitem[{Newman {et~al.}(1968)Newman, Cooper, Davis, Faber, Coil, Guhathakurta, 
  Koo, Phillips, Conroy, Dutton, Finkbeiner, Gerke, Rosario, Weiner, Willmer, 
  Yan, Harker, Kassin, Konidaris, Lai, Madgwick, Noeske, Wirth, Connolly, Kaiser, 
  Kirby, Lemaux, Lin, Lotz, Luppino, Marinoni, Matthews, Metevier, \& Schiavon}]{Newman2012}
Newman, J.~M., Cooper, M.~C., Davis, M., {et~al.} 2012, eprint arXiv:1203.3192

\bibitem[{Peng {et~al.}(2010)Peng, Lilly, Kova\v{c}, Bolzonella, Pozzetti,
  Renzini, Zamorani, Ilbert, Knobel, Iovino, Maier, Cucciati, Tasca, Carollo,
  Silverman, Kampczyk, de~Ravel, Sanders, Scoville, Contini, Mainieri,
  Scodeggio, Kneib, {Le F\`{e}vre}, Bardelli, Bongiorno, Caputi, Coppa, de~la
  Torre, Franzetti, Garilli, Lamareille, {Le Borgne}, {Le Brun}, Mignoli,
  Montero, Pello, Ricciardelli, Tanaka, Tresse, Vergani, Welikala, Zucca,
  Oesch, Abbas, Barnes, Bordoloi, Bottini, Cappi, Cassata, Cimatti, Fumana,
  Hasinger, Koekemoer, Leauthaud, Maccagni, Marinoni, McCracken, Memeo, Meneux,
  Nair, Porciani, Presotto, \& Scaramella}]{Peng2010}
Peng, Y.-j., Lilly, S.~J., Kova\v{c}, K., {et~al.} 2010, \apj, 721, 193

\bibitem[{P\'{e}rez‐Gonz\'{a}lez {et~al.}(2008)P\'{e}rez‐Gonz\'{a}lez,
  Trujillo, Barro, Gallego, Zamorano, \& Conselice}]{PerezGonzalez2008}
P\'{e}rez‐Gonz\'{a}lez, P.~G., Trujillo, I., Barro, G., {et~al.} 2008, \apj,
  687, 50

\bibitem[{Postman \& Geller(1984)}]{Postman1984}
Postman, M., \& Geller, M.~J. 1984, \apj, 281, 95

\bibitem[{Prochaska {et~al.}(2007)Prochaska, Rose, Caldwell, Castilho,
  Concannon, Harding, Morrison, \& Schiavon}]{Prochaska2007}
Prochaska, L.~C., Rose, J.~A., Caldwell, N., {et~al.} 2007, \aj, 134, 321

\bibitem[{Salim {et~al.}(2005)Salim, Charlot, Rich, Kauffmann, Heckman, Barlow,
  Bianchi, Byun, Donas, Forster, Friedman, Jelinsky, Lee, Madore, Malina,
  Martin, Milliard, Morrissey, Neff, Schiminovich, Seibert, Siegmund, Small,
  Szalay, Welsh, \& Wyder}]{Salim2005}
Salim, S., Charlot, S., Rich, R.~M., {et~al.} 2005, \apj, 619, L39

\bibitem[{Salim {et~al.}(2009)Salim, Dickinson, {Michael Rich}, Charlot, Lee,
  Schiminovich, P\'{e}rez-Gonz\'{a}lez, Ashby, Papovich, Faber, Ivison, Frayer,
  Walton, Weiner, Chary, Bundy, Noeske, \& Koekemoer}]{Salim2009}
Salim, S., Dickinson, M., {Michael Rich}, R., {et~al.} 2009, \apj, 700, 161

\bibitem[{Schawinski {et~al.}(2009)Schawinski, Virani, Simmons, Urry, Treister,
  Kaviraj, \& Kushkuley}]{Schawinski2009}
Schawinski, K., Virani, S., Simmons, B., {et~al.} 2009, \apj, 692, L19

\bibitem[{Schiminovich {et~al.}(2010)Schiminovich, Catinella, Kauffmann,
  Fabello, Wang, Hummels, Lemonias, Moran, Wu, Giovanelli, Haynes, Heckman,
  Basu-Zych, Blanton, Brinchmann, Budav\'{a}ri, Gon\c{c}alves, Johnson,
  Kennicutt, Madore, Martin, Rich, Tacconi, Thilker, Wild, \&
  Wyder}]{Schiminovich2010}
Schiminovich, D., Catinella, B., Kauffmann, G., {et~al.} 2010, \mnras, 408, 919

\bibitem[{Schmidt}(1968)]{Schmidt1968}
Schmidt, M. 1968, \apj, 151, 393

\bibitem[{Strateva {et~al.}(2001)Strateva, Ivezi\'{c}, Knapp, Narayanan,
  Strauss, Gunn, Lupton, Schlegel, Bahcall, Brinkmann, Brunner, Budav\'{a}ri,
  Csabai, Castander, Doi, Fukugita, Győry, Hamabe, Hennessy, Ichikawa, Kunszt,
  Lamb, McKay, Okamura, Racusin, Sekiguchi, Schneider, Shimasaku, \&
  York}]{Strateva2001}
Strateva, I., Ivezi\'{c}, v., Knapp, G.~R., {et~al.} 2001, \aj, 122, 1861

\bibitem[{Takamiya {et~al.}(1995)Takamiya, Kron, \& Kron}]{Takamiya1995}
Takamiya, M., Kron, R.~G., \& Kron, G.~E. 1995, \aj, 110, 1083

\bibitem[{Thilker {et~al.}(2010)Thilker, Bianchi, Schiminovich, {Gil de Paz},
  Seibert, Madore, Wyder, Rich, Yi, Barlow, Conrow, Forster, Friedman, Martin,
  Morrissey, Neff, \& Small}]{Thilker2010}
Thilker, D.~A., Bianchi, L., Schiminovich, D., {et~al.} 2010, \apj, 714, L171

\bibitem[{Ueda {et~al.}(2003)Ueda, Akiyama, Ohta, \& Miyaji}]{Ueda2003}
Ueda, Y., Akiyama, M., Ohta, K., \& Miyaji, T. 2003, \apj, 598, 886

\bibitem[{van~den Bergh(2002)}]{VandenBergh2002}
van~den Bergh, S. 2002, PASP, 114, 797

\bibitem[{Visvanathan \& Sandage(1977)}]{Visvanathan1977}
Visvanathan, N., \& Sandage, A. 1977, \apj, 216, 214

\bibitem[{Weiner {et~al.}(2005)Weiner, Phillips, Faber, Willmer, Vogt, Simard,
  Gebhardt, Im, Koo, Sarajedini, Wu, Forbes, Gronwall, Groth, Illingworth,
  Kron, Rhodes, Szalay, \& Takamiya}]{Weiner2005}
Weiner, B.~J., Phillips, A.~C., Faber, S.~M., {et~al.} 2005, \apj, 620, 595

\bibitem[{Willmer {et~al.}(2006)Willmer, Faber, Koo, Weiner, Newman, Coil,
  Connolly, Conroy, Cooper, Davis, Finkbeiner, Gerke, Guhathakurta, Harker,
  Kaiser, Kassin, Konidaris, Lin, Luppino, Madgwick, Noeske, Phillips, \&
  Yan}]{Willmer2006}
Willmer, C. N.~A., Faber, S.~M., Koo, D.~C., {et~al.} 2006, \apj, 647, 853

\bibitem[{Wyder {et~al.}(2007)Wyder, Martin, Schiminovich, Seibert, Budavari,
  Treyer, Barlow, Forster, Friedman, Morrissey, Neff, Small, Bianchi, Donas,
  Heckman, Lee, Madore, Milliard, Rich, Szalay, Welsh, \& Yi}]{Wyder2007}
Wyder, T.~K., Martin, D.~C., Schiminovich, D., {et~al.} 2007, \apjs, 173, 293

\end{thebibliography}
\end{document}